\documentclass[a4paper]{panl}
\usepackage{cite}
\usepackage{wrapfig}
\usepackage{graphicx}
\usepackage{amssymb}
\usepackage{amsfonts}
\usepackage{amsmath}
\usepackage{longtable}
\usepackage{rotating}
\usepackage{lscape}
\usepackage{epsfig}
\usepackage{multirow}
\usepackage{hyperref}

\originalTeX
\begin{document}
\issuearea{Physics of Elementary Particles and Atomic Nuclei. Theory}

\title{Towards Lagrangian dynamics for constrained   mixed-symmetric  interacting  higher-spin fields
} \maketitle \setcounter{footnote}{0} \authors{A.A.\,Reshetnyak$^{a,b}$\footnote{E-mail: reshet@tspu.ru}}
\from{$^{a}$\, Center for Theoretical Physics, Tomsk State
Pedagogical University, 634061, Tomsk, Russia}
\from{$^{b}$\,National
Research Tomsk Polytechnic University, 634050, Tomsk, Russia}
\begin{abstract}

\vspace{0.2cm}

The necessary and sufficient conditions  to construct consistent Lagran\-gi\-an formulation for irreducible   interacting massless  higher-spin (HS) fields on $d$-dimen\-sio\-nal Minkowski space within approach  with incomplete  BRST operator  and off-shell   holonomic constraints are found. It is shown that in addition to superconmmuting of incomplete BRST operator with appropriate traceless and Young constraints, which annihilate the field and gauge parameter vectors, these constraints  should form Abelian superalgebra both with  BRST operator and with operators of cubic, quartic and etc. vertices.
The consistent deformation of free model with constrained HS fields with integer spin  requires for the cubic vertex to be by BRST-closed, traceless and Young-symmetric solution of the generating equations. The explicit form for the vertices for irreducible constrained interacting fields are obtained by means of  projectors on traceless and Young-symmetric modes.

\end{abstract}
\vspace*{6pt}


\label{sec:intro}
\section*{Introduction}

 Theory of interacting HS fields as the part of theoretical
 high-energy physics  deals with   modern concepts  (for a review see, e.g.,
 \cite{reviews3},  \cite{reviewsV}) devoting for searching  new ways  for the unification of fundamental interaction of elementary particles beyond
the Standard Model. A connection with  (Super)string  Field Theory permits one to include  massless fields of  spins $s>2$ in  HS Gravity (see \cite{Snowmass} and references therein)  to state the properties of yet unknown consistent theory of quantum gravity. Interacting massive and massless HS fields in
constant-curvature spaces  give alternative    possible
insight into the origin of Dark Matter and Dark Energy,   beyond the
 models with  sterile neutrinos  and vector
massive fields  being by probable  candidates for Dark
Matter, see  for reviews  \cite{DM1}, \cite{Odintsov1}.
Conventional application of the  Feynman diagrammatic techniques in covariant quantization methods for the models with interacting HS fields is usually realized within Lagrangian formulations (LFs) starting from  ones for free  HS fields and then undergoing some deformation procedure.

There exists two  ways to construct gauge-invariant   Lagrangians for HS fields both in metric- and in frame-like formalisms. In the  first one (developed, e.g. in \cite{PT}, \cite{BPT}, \cite{BR}), the fields remains by unconstrained, i.e. all the conditions (d'Alamberti\-an or Dirac, divergentless; ($\gamma$-)traceless, mixed(Young)-symmetric), which select the irreducible Poincare group representation  with given mass and spin are equally included  to get Lagrangian. In the second one (see e.g.  \cite{AlkalTipGrigoriev}, \cite{Reshetnyak_con}), part from them, usually related with traces and Young symmetries, are consistently imposed  on the fields and gauge parameters as additional constraints. On the level of free and interacting fields there exist two effective approaches to these aims, known as the  BRST   aprroaches respectively with complete, $Q$, and  incomplete, $Q_c$,  BRST operators which make LFs for the same HS field in $d$-dimensional Minkowski space-time was shown to be equivalent in \cite{Reshetnyak_con}. For the cubic interaction case (see \cite{Manvelyan},  \cite{Joung}, \cite{frame-like1}, \cite{Metsaev0512}, \cite{Metsaev0712}, \cite{BRST-BV3}, \cite{frame-like2}, \cite{BKTW}  in various methods) the respective classification for irreducible HS fields was given in light-cone formalism by Metsaev \cite{Metsaev0512}. The covariant form  of this result within constrained BRST approach \cite{BRST-BV3} works perfectly for reducible interacting  HS fields, whereas for irreducible ones meets the obstacle due to  not taking into account the influence of holonomic constraints with traces and Young symmetries  on the structure of the vertices. That fact destroys the traceless and
mixed-symmetric properties both for the deformed  equations of motion and gauge transformations, thus leading to another number of physical degrees of freedom (p.d.f.) (one of independent initial data)  as compared with  case of free HS fields. In our papers \cite{BRcub}, \cite{Rcubmasless},  \cite{BRcubmass} (see, for review  \cite{Rrev} and \cite{BKStwis} for  $d=4$ case)  the solutions for Lagrangian cubic vertices  has been derived for interacting unconstrained
integer massless and massive totally-symmetric  fields in $\mathbb{R}^{1,d-1}$ within $Q$-BRST approach (developed, for example, in \cite{reviews3}, \cite{BPT},  \cite{BR}, \cite{BKR}). The solutions have additional terms as compared to \cite{BRST-BV3}, automatically  preserves the number of p.d.f. when passing to interacting case, thus providing for the first time construction of the covariant cubic vertices for irreducible HS fields.
The necessary and sufficient conditions for application of constrained BRST approach for construction interacting Lagrangians for irreducible
totally-symmetric HS fields on flat space-times were recently formulated in \cite{necessBR}. However, for the models which work with constrained interacting mixed-symmetric irreducible HS fields this problem still remains by open one.

The first  main purpose of the paper is to develop the consistent  deformation procedure for the approach with incomplete, $Q_c$, BRST operator  to get cubic, quartic  and so on vertices for LFs for the fields above obtained from  methods with complete and incomplete BRST operators.

The paper is organized as follows.  In
Section~\ref{sec:preparation}, the results of the BRST construction
using a incomplete BRST operator are presented. In
Section~\ref{sec:consistent}, we derive necessary and sufficient
conditions for the incomplete BRST operator, BRST-extended holonomic
constraints, vertex operators to have non-contradictory Lagrangian
dynamics.  Conclusion resumes the results.

We use the conventions of \cite{BRcub}, \cite{BRcubmass} : $\eta_{\mu\nu} = diag (+, -,...,-)$ for a metric tensor with Lorentz indices $\mu, \nu = 0 ,1,...,d-1$,
the notation $\epsilon(F)$,$gh(F)$, $[H,\,G\}$, $[y]$, $\vec{s}_{k}$, $\theta_{m,l}$
 for the Grassmann parity and ghost number of a homogeneous quantity $F$,
as well as the supercommutator of quantities $H,G$, the integer part of a number $y$, and the integer-valued vector of generalized spin $(s_{1},s_{2},..., {s}_{k})$.  Heaviside
$\theta$-symbol to be equal to $1(0)$ when $m>l(m\leq l)$.

\section{Incomplete BRST operator for free mixed-symmetric HS fields}
\label{sec:preparation}

The unitary irreducible representations of
Poincar\'e  group  with generalized integer spin $\vec{s}_{k}$ can be realized
using the $\mathbb{R}$-valued mixed-symmetric tensor fields
$\phi^{\mu^1_1...\mu^1_{s_{1}},...,\mu^k_1...\mu^k_{s_k}}(x) \equiv \phi_0^{\mu^1(s_1)...\mu^k(s_k)}$ for $k\leq [(d-2+\theta_{m,0})/2]$, included into
basic vector $|\phi\rangle$ subject to the conditions
\begin{eqnarray}\label{irrepint}
     \hspace{-1.3em}  &\hspace{-1.3em}& \hspace{-1.3em}      \big(g^i_0 -d/2; l_0,\, l_i,\, l_{ij}, t_{rs}\big)|\phi\rangle  = (s_i;\vec{0})|\phi\rangle, \ |\phi\rangle  =  \sum_{s_1=0}^{\infty}\sum_{s_2=0}^{s_1}\cdots\sum_{s_k=0}^{s_{k-1}}
\frac{\imath^{\sum_is_i}}{s_1!... s_k!} \\
&  &  \label{comrels}\times
  \phi_{\mu^1(s_1)...\mu^k(s_k)}\,
\prod_{i=1}^k\prod_{l_i=1}^{s_i} a^{+\mu^i_{l_i}}_i|0\rangle, \quad  [a^i_{\mu^i}, a_{\nu^j}^{j+}]=-\eta_{\mu^i\nu^j}\delta^{ij},
\end{eqnarray}
(for $i\leq j$ and $r<s$). The operators  of number particles $g^i_0$, d'Alambert  $l_0,\,$
divergent $l_i,$ traceless $l_{ij}$, Young-symmetry $t_{rs}$ constraints above  are defined in the Fock space $\mathcal{H}$
with the Grassmann-even oscillators $a^i_{\mu^i}, a_{\nu^j}^{j+}$,   as follows
\begin{eqnarray}\label{FVoper}
&\hspace{-0.7em} &\hspace{-0.7em}  \big( g^i_0,\, l_0,\, l_i,\, l_{ij}, t_{rs}\big) = \big( -\frac{1}{2}\big\{a^{+i}_{\mu},\, a^{i\mu}\big\},\,\partial^\nu\partial_\nu+ m^2 ,\, - \imath a^\nu_i  \partial_\nu ,\, \frac{1}{2}a^\mu_i a_{j\mu}, a^{\mu+}_r a_{s\mu}\big).\nonumber
\end{eqnarray}
The free dynamics of the field with
definite spin $\vec{s}_k$ in the framework of constrained  BRST approach  is described by the $(k-1)$-th stage reducible gauge theory with the gauge invariant action given on the configuration space $M^{\vec{s}_k}_{c|cl}$ which includes in addition to $\phi_0^{\mu^1(s_1)...\mu^k(s_k)} $ maximally  the set of auxiliary  fields by number $\sum_{e=1}^{[k/2]}C^e_kC^e_{k+1}$, for $ C^e_k = k!/(e! (k-e)!)$ (for massive case a family of this set of fields) incorporated into the vector $|\chi_c\rangle_{\vec{s}_k}$
\begin{eqnarray}
\label{PhysStatetot} \mathcal{S}^m_{0|\vec{s}_k}[\phi_i]=
\mathcal{S}^m_{0|\vec{s}_k}[|\chi_c\rangle] = \hspace{-0.3em} \int\hspace{-0.2em} d\eta_0 {}_{\vec{s}_k}\langle\chi_c|
Q_c|\chi_c\rangle_{\vec{s}_k}, \\\label{PhysStatetot1}
 \delta\big(|\chi_c\rangle_{\vec{s}_k},\, |\Lambda^e_c\rangle_{\vec{s}_k}\big)  =  Q_c|\big(\Lambda^0_c\rangle_{\vec{s}_k} , \theta_{e,k-1}|\Lambda^{e+1}_c\rangle_{\vec{s}_k}\big)
\end{eqnarray}
(for $e=0,1,...,k-1$) and subject to the additional off-shell constraints
\begin{eqnarray}\label{L11T}
    && L_{ij} \big(|\chi_c\rangle_{\vec{s}_k}, |\Lambda^e_c\rangle_{\vec{s}_k}\big) = T_{rs} \big(|\chi_c\rangle_{\vec{s}_k}, |\Lambda^e_c\rangle_{\vec{s}_k}\big)=0, \ \ e=0,...,k-1
\end{eqnarray}
Here $\eta_0$, $Q_c$, $L_{ij}$, $T_{rs}$  and $ |\Lambda^e_c\rangle_s$  be respectively a zero-mode ghost, nilpotent incomplete BRST operator, BRST-extended  traceless, Young-symmetry constraints (corresponding to Labastida idea \cite{Labastida}) and $e$-th level  parameter of reducible  gauge transformations:
\begin{eqnarray}
&\hspace{-0.7em} &\hspace{-0.7em}  {Q}_c  =
\eta_0l_0+\sum_i\big(\eta_i^+\check{l}_i+\check{l}_i^{+}\eta_i
 + {\imath}\eta_i^+\eta_i{\cal{}P}_0\big),
\label{Qctotsym}\\
\hspace{-1.5em}&\hspace{-1.53em} &\hspace{-1.5em}  \big(\hspace{-0.1em} L_{ij},\hspace{-0.1em} T_{rs} \hspace{-0.1em}\big) \hspace{-0.15em} =  \big(\hspace{-0.15em}
 l_{ij}\hspace{-0.1em}-\hspace{-0.1em}\textstyle\frac{\theta_{m,0}}{2}d_id_j\hspace{-0.1em}+\hspace{-0.1em}\frac{1}{2}\big(\hspace{-0.1em}\eta_{i} \mathcal{P}_{j}\hspace{-0.1em}+\hspace{-0.1em}\eta_{j} \mathcal{P}_{i}\hspace{-0.1em}\big),\, t_{rs}\hspace{-0.1em}-\hspace{-0.1em}\theta_{m,0}d^+_rd_s\hspace{-0.1em} - \hspace{-0.1em}\eta_{r}^+ \mathcal{P}_{s}\hspace{-0.1em}-\hspace{-0.1em}\mathcal{P}^+_{r}\eta_{s}\hspace{-0.1em} \big)
 \label{Qctotsym1}
\end{eqnarray}
(for $(\check{l}_i, \check{l}_i^{+}) = ({l}_i+md_i, {l}_i^{+}+md^+_i) $ with additional Grassmann-even oscillators $d_i, d_i^+$:  $[d_i, d_j^+]$ $=\delta_{ij}$).
The Grassmann-odd  ghost oscillators $\eta_0,  \mathcal{P}_{0}$,  $\eta_i,  \mathcal{P}_{i}^+$, $\eta_i^+$,  $\mathcal{P}_{i}$ correspond to the system of first-class differential constraints $l_0, \check{l}_i,\check{l}_i^+$ with algebra $[\check{l}_i,\, \check{l}_j^+]=\delta_{ij} l_0$  and satisfy to the non-vanishing anticommutators
    \begin{equation}\label{ghanticomm}
  \{\eta_0, \mathcal{P}_0\}= \imath,\ \  \{\eta_i, \mathcal{P}_j^+\}=\delta_{ij},  \ \ (\epsilon, gh)\eta_{...} = (\epsilon,  - gh)\mathcal{P}_{...}=(1,1).
\end{equation}
The label "$\vec{s}_k$" at field and gauge parameter vectors
\begin{eqnarray}
\label{extconstsp5}
 &\hspace{-0.5em} &\hspace{-0.5em}  |\chi_c\rangle_{\vec{s}_k} = |\Phi_0\rangle_{\vec{s}_k}- \sum_i\mathcal{P}_i^{+}\big(\eta_0|\Phi_{0i}\rangle_{\vec{s}_k-\delta_{ki}}+\sum_j\eta_j^{+}|\Phi_{ij}\rangle_{\vec{s}_k-\delta_{kj}-\delta_{ki}}\big)\\
&\hspace{-0.5em} &\hspace{-0.5em} \ +\sum_{h>1}^{[k/2]}\sum_{i_1>...>i_h}\prod_{p=1}^h\hspace{-0.2em}\mathcal{P}_{i_p}^{+}\Big(\eta_0\hspace{-0.5em}\sum_{j_1>...>j_{h-1}}\prod_{r=1}^{h-1}
\hspace{-0.2em}\eta_{j_r}^+
|\Phi_{0(i)_h(j)_{h-1}}\rangle_{\big(\hspace{-0.1em}\vec{s}_k-\sum_{p=1}^h\delta_{ki_p}-\sum_{p=1}^{h-1}\delta_{kj_p}\hspace{-0.1em}\big)}\nonumber\\
&\hspace{-0.5em} &\hspace{-0.5em} \quad
+\sum_{j_1>...>j_h}\prod_{r=1}^h
\eta_{j_r}^+
|\Phi_{(i)_h(j)_{h}}\rangle_{\big(\vec{s}_k-\sum_{p=1}^h\delta_{ki_p}-\sum_{p=1}^{h}\delta_{kj_p}\big)}\Big), \nonumber\\
&\hspace{-0.5em} &\hspace{-0.5em}\label{extconstsp7} |\Lambda^e_c\rangle_{\vec{s}_k} = \sum_{i_1>...>i_e}\prod_{p=1}^e\mathcal{P}_{i_p}^{+}|\Xi^e_{(i)_e}\rangle_{\big(\vec{s}_k-\sum_{p=1}^e\delta_{ki_p}\big)}+\sum_{h=1}^{[k-e/2]}\sum_{i_1>...>i_h}\prod_{p=1}^h\mathcal{P}_{i_p}^{+}
\nonumber\\
&\hspace{-0.5em} &\hspace{-0.5em}
\quad
\times \Big(\eta_0 \sum_{j_{e+1}>...>j_{h-1}}
\prod_{r=1}^{h-e-1}
\eta_{j_r}^+|\Xi^e_{0(i)_h(j)_{h-e-1}}\rangle_{\big(\vec{s}_k  -\sum_{p=1}^h\delta_{ki_p}-\sum_{r=1}^{h-e-1}\delta_{kj_r}\big)}
\nonumber\\
&\hspace{-0.5em} &\hspace{-0.5em}
\quad  +\sum_{j_{e+1}>...>j_h}\prod_{r=1}^{h-e}
\eta_{j_r}^+
|\Xi^e_{(i)_h(j)_{h-e}}\rangle_{\big(\vec{s}_k-\sum_{p=1}^h\delta_{ki_p}-\sum_{r=1}^{h-e}\delta_{kj_r}\big)}\Big), \nonumber\\
 &\hspace{-0.5em} &\hspace{-0.5em}  |\Phi_{(i)_h(j)_{h}}\rangle_{\vec{s}_k-...}\hspace{-0.1em}=\hspace{-0.1em}\sum_{p=1}^k \sum_{l_p=0}^{s_p-...}\prod_{r=1}^k\frac{(d^+_r)^{l_r}}{l_r!}|\phi_{{(i)_h(j)_{h}}|(l)_k}(a^+)\rangle_{\vec{s}_k-\vec{l}_k-...},     \end{eqnarray}
(for $|\phi_{(0)|(0)}\rangle_{\vec{s}_k}\equiv |\phi\rangle_{\vec{s}_k}$) means that these vectors are proper eigen-vectors for the incomplete spin operator with definite spin value $\vec{s}_k$
\begin{eqnarray}
&&   \sigma^i_{c} \Big(|\chi_c\rangle_{\vec{s}_k}, |\Lambda^e_c\rangle_{\vec{s}_k}\Big) = \Big(s_i-1+\frac{d + \theta_{m,0}}{2}\Big)\Big(|\chi_c\rangle_{\vec{s}_k}, |\Lambda^e_c\rangle_{\vec{s}_k}\Big), \label{s11} \\
 &&  {\sigma}^i_c   =   g^i_0 + \theta_{m,0}\big( d_i^{+}d_i^{} + \frac{1}{2}\big)+ \sum_i\big(\eta_i^{+}\mathcal{P}_{i}
-\eta_i\mathcal{P}_{i}^{+}\big).
\end{eqnarray}
The incomplete
 BRST operator $Q_c$ forms with system of  holonomic  constraints and incomplete spin operator closed superalgebra \cite{Reshetnyak_con}:
   \begin{eqnarray} \label{QsL11}
 &&   (Q_c)^2 \ = \  [ Q_c, \,  L_{ij}\} \ =\ [ Q_c, \,  T_{rs}\} \ =\ [ Q_c, \,  {\sigma}^l_{c}\}= 0, \\
  && \ [L^{ij},\,\sigma^l_{c}\}= \delta^{l\{i} L^{j\}i},\ \ \  [T^{rs},\,\sigma^l_{c}\}\ =\  \delta^{ls} T^{rl}-\delta^{rs} T^{sl} . \nonumber
\end{eqnarray}
Note, that both the equations of motion, $Q_c|\chi_c\rangle_{\vec{s}_k} =0$, and,  that any field representative ($\big|\widetilde{\chi}_c\rangle_{\vec{s}_k}$) from the gauge orbit
\begin{equation}\label{goc}
\mathcal{O}_{0|\chi_c} = \big\{\big|\widetilde{\chi}_c\rangle_{\vec{s}_k} \big|\, \  \big|\widetilde{\chi}_c\rangle_{\vec{s}_k} =\big|{\chi}_c\rangle_{\vec{s}_k} +  Q_c \big|\Lambda^0_c\rangle_{\vec{s}_k}, \, \forall \big|\Lambda^0_c\rangle_{\vec{s}_k}\big\}
\end{equation}
 remains by traceless and mixed-symmetric if the field $\big|{\chi}_c\rangle$ and  gauge parameter $ \big|\Lambda^0_c\rangle$  are traceless and mixed-symmetric because of the commutation of
 $L_{ij}$, $T_{rs}$  with $Q_c$. The same property is true for any representative ($\big|\widetilde{\Lambda}^e_c\rangle_{\vec{s}_k}$) from $e$-th level gauge parameter orbit $\mathcal{O}_{0|\Lambda^e_c}$.

One can show that after resolving the algebraic constraints
and eliminating the auxiliary fields from equations of motion, the
theory under consideration is reduced to Labastida  \cite{Labastida} (for massive case, generalization of Singh-Hagen Lagrangian) form in terms of
mixed-symmetric double traceless tensor field $\phi_0^{\mu^1(s_1)...\mu^k(s_k)}$  and mixed-symmetric
traceless gauge parameters.

\section{Consistent deformation procedure  for interacting  higher-spin fields}
\label{sec:consistent}

To  include the  interaction   we introduce $p$, $p\geq 3$, copies of LFs (to adapt the model for Yang-Mills type interactions with gauge group $SU(N)$, for $p=N^2-1$) with vectors
$|\chi^{(t)}_c\rangle_{(\vec{s}_k)_t}$, reducible gauge parameters
$|\Lambda^{(t)e}_c\rangle_{\vec{s}_{k_t}}$,
corresponding vacuum vectors $|0\rangle^t$ and oscillators for
$t=1,...,p$ (with notation $(\vec{s}_k)_t\equiv  (\vec{s}_{k_1}, ..., \vec{s}_{k_t})$ for different in general values at $k_1$,..., $k_t$). It permits to  define the deformed  action and gauge transformations up to $r$-tic vertices, $r=3,4,...,e$  in  powers of deformation  constant $g$, starting from sum of $p$ copies of LFs for free   HS fields and then from cubic, quartic and so on vertices:
\vspace{-1ex}
\begin{eqnarray}\label{S[e]}
  && S^{(m)_p}_{[e]|(\vec{s}_k)_p}[(\chi_c)_p] \  = \  \sum\nolimits_{t=1}^{p} \mathcal{S}^{m_t}_{0|\vec{s}_{k_t}}[\chi^{(t)}_c]   + \sum\nolimits_{h=1}^e g^h S^{(m)_p}_{h|(\vec{s}_k)_p}[(\chi_c)_p],
  \end{eqnarray}
\vspace{-1ex}  where
\begin{eqnarray}\label{S[3]}
\hspace{-0.7em}&\hspace{-0.7em}&\hspace{-0.7em}  S^{(m)_p}_{1|(\vec{s}_k)_p}[(\chi_c)_p] =   \sum_{1\leq i_1<i_2<i_3\leq p} \hspace{-1.0em} \int \prod_{j=1}^3 d\eta^{(i_j)}_0  \Big( {}_{\vec{s}_{k_{i_j}}}\langle \chi^{(i_j)}_c
  \big|  V^{(3)}_c\rangle^{(m)_{(i)_3}}_{(\vec{s}_k)_{(i)_3}}+h.c. \Big)  , \\
  \label{S[4]}
\hspace{-0.9em}&\hspace{-0.9em}&\hspace{-0.9em}  S^{(m)_p}_{2|(\vec{s}_k)_p}[(\chi_c)_p] =   \sum_{1\leq i_1<i_2<i_3<i_4\leq p}\hspace{-1.0em}  \int \prod_{j=1}^4 d\eta^{(i_j)}_0  \Big( {}_{\vec{s}_{k_{i_j}}}\langle \chi^{(i_j)}_c
  \big|  V^{(4)}_c\rangle^{(m)_{(i)_4}}_{(\vec{s}_k)_{(i)_4}}+h.c. \Big)  , \\
   && \ldots \ \ldots\ \ldots \ \ldots\ \ldots \ \ldots\ \ldots \ \ldots\ \ldots \ \ldots \ \ldots \ \ldots\ \ldots \ \ldots \nonumber\\
    \label{S[+]}
\hspace{-0.9em}&\hspace{-0.9em}&\hspace{-0.9em}  S^{(m)_p}_{e|(\vec{s}_k)_p}[(\chi_c)_p] =   \sum_{1\leq i_1<i_2<...<i_e\leq p}  \hspace{-1.0em}\int \prod_{j=1}^e d\eta^{(i_j)}_0  \Big( {}_{\vec{s}_{k_{i_j}}}\langle \chi^{(i_j)}_c
  \big|  V^{(e)}_c\rangle^{(m)_{(i)_e}}_{(\vec{s}_k)_{(i)_e}}+h.c. \Big) ,
\end{eqnarray}
also for deformed $l$-th level gauge transformations: $\delta^l_{[e]}  |\Lambda^{(t)l}_c\rangle_{\vec{s}_{k_t}}=(\delta^l_0+\sum_{q=1}^eg^q\delta^l_q )|\Lambda^{(t)l}_c\rangle_{\vec{s}_{k_t}} $ (for $ |\Lambda^{(t)-1}_c\rangle_{\vec{s}_{k_t}}\equiv |\chi^{(t)}_c\rangle_{\vec{s}_{k_t}}$):
\begin{eqnarray}\label{gt1}
\hspace{-1.1em}&\hspace{-1.1em}&\hspace{-1.1em} \delta^l_1|\Lambda^{(t)l}_c\rangle_{\vec{s}_{k_t}}\hspace{-0.1em} = \hspace{-0.1em} - \hspace{-1.0em}\sum_{1\leq i_1<i_2\leq p}\hspace{-0.5em} \int \hspace{-0.3em}\prod_{j=1}^2 d\eta^{(i_j)}_0 \hspace{-0.15em} \Big[\hspace{-0.1em} {}_{\vec{s}_{k_{\{i_1}}}\hspace{-0.25em}\langle \chi^{(\{i_1)}_c
  \big|  {}_{\vec{s}_{k_{i_2\}}}}\hspace{-0.25em}\langle \Lambda^{(i_2\})l+1}_c
  \big| {V}{}^{(3)l}_c\rangle^{(m)_{(i)_2t}}_{(\vec{s}_k)_{(i)_2{}t}}\hspace{-0.1em}+h.c. \hspace{-0.15em}\Big]\hspace{-0.1em}  ,
  \end{eqnarray}
 \begin{eqnarray}
   &\hspace{-0.5em}&\hspace{-0.5em} \ldots \ \ldots\ \ldots \ \ldots\ \ldots \ \ldots\ \ldots \ \ldots\ \ldots \ \ldots \ \ldots \ \ldots\ \ldots \ \ldots \nonumber\\
    \label{gte}
&\hspace{-0.5em}&\hspace{-0.5em}  \delta^l_e|\Lambda^{(t)l}_c\rangle_{\vec{s}_{k_t}} =   -\sum_{1\leq i_1<...<i_{e-1}\leq p} \hspace{-1.0em} \int \prod_{j=1}^{e-1} d\eta^{(i_j)}_0  \Big[{}_{\vec{s}_{k_{i_1}}}\langle \chi^{(\{i_1)}_c
  \big| \ldots {}_{\vec{s}_{k_{i_{e-2}}}}\langle \chi^{(i_{e-2})}_c
  \big| \\
  &\hspace{-0.5em}&\hspace{-0.5em} \phantom{\delta_2|\chi^{(j)}_c\rangle_{s_j}\ \ }\otimes
   {}_{\vec{s}_{k_{i_{e-1}}}}\langle\Lambda^{(i_{e-1}\})l+1}_c
  \big|  {V}{}^{(e)l}_c\rangle^{(m)_{(i)_{e-1}t}}_{(\vec{s}_k)_{(i)_{e-1}t}}+h.c. \Big] . \nonumber
\end{eqnarray}
Here we have used the notations $(\chi_c)_p= (\chi^{(1)}_c, \chi^{(2)}_c, ..., \chi^{(p)}_c)$, the symmetrization  of indices $\{i_1,...,i_{e-1}\}$  with $r$-tic vertices $\big|  {V}{}^{(r)l}_c\rangle$, for $l=-1,0,...,k-1$.
The preservation for the interacting theory constructed from  initial  actions $\mathcal{S}^{m_t}_{0|\vec{s}_{k_t}}$, $t=1,...,p$  the number $N_t$  of p.d.f.  determined by LFs for free HS field  with spin $\vec{s}_{k_t}$, requires  that the sum of all p.d.f. would be unchangeable, i.e. $\sum_{t}N_t= \mathrm{const}$. The property will be guaranteed, first, if the deformed  ac\-tion $S^{(m)_p}_{[e]|(\vec{s}_k)_{p}}$ will satisfy to sequence of new Noether identities in powers of $g$:
\begin{eqnarray}
       &g^1:& \delta_0 S^{(m)_p}_{1|(\vec{s}_k)_p}[(\chi_c)_p]   + \delta_1 \mathcal{S}^{(m)_p}_{[0]|(\vec{s}_k)_p}[(\chi_c)_p]  =0   , \label{g1} \\
    &g^2:& \delta_0 S^{(m)_p}_{2|(\vec{s}_k)_p}[(\chi_c)_p]+ \delta_1 S^{(m)_p}_{1|(\vec{s}_k)_p}[(\chi_c)_p] + \delta_2 \mathcal{S}^{(m)_p}_{0|(\vec{s}_k)_p}[(\chi_c)_p] = 0 , \label{g2} \\
&& \ldots \ \ldots \  \ldots \  \ldots \ \ldots \ \ldots \ \ldots \ \ldots \ \ldots \ \ldots \ \nonumber \\
    &g^{e}:& \sum_{j=0}^e \delta_j S^{(m)_p}_{e-j|(\vec{s}_k)_p}[(\chi_c)_p] =0 , \label{ge}
 \end{eqnarray}
 second, from $\delta^{l}_{[\infty]}\delta^{l-1}_{[\infty]}|\Lambda^{(t)l-1}_c\rangle_{\vec{s}_{k_t}}\vert_{\partial S^{(m)_p}_{[\infty]|(\vec{s}_k)_p}=0} =0 $ for all  levels of
 gauge transformations:
 \begin{eqnarray}
{g^1}: && \Big(\delta^{l}_1\delta^{l-1}_0+ \delta^{{l}}_0\delta^{l-1}_1 \Big)|\Lambda^{(t)l-1}_c\rangle_{\vec{s}_{k_t}}\vert_{\partial S^{(m)_p}_{[1]}=0} =0 , \label{g0Ltttt1}  \\
{g^2}: && \Big(\delta^{l}_2\delta^{l-1}_0+ \delta^{{l}}_1\delta^{l-1}_1+\delta^{l}_0\delta^{l-1}_0 \Big)|\Lambda^{(t)l-1}_c\rangle_{\vec{s}_{k_t}} \vert_{\partial S^{(m)_p}_{[2]}=0} =0,\label{g0Ltttt2}\\
\vspace{-0.5ex}&& \ldots\ldots\ldots\ldots\ldots\ldots\ldots\ldots\ldots \nonumber \\
\vspace{-0.5ex} {g^e}:  &&  \Big(\delta^{l}_e\delta^{l-1}_0 + \textstyle\sum_{p=1}^e\delta^l_{e-p}\delta^{l-1}_p  \Big)|\Lambda^{(t)l-1}_c\rangle_{\vec{s}_{k_t}} \vert_{\partial S^{(m)_p}_{[e]}=0} =0 \label{g0Ltttte}
\end{eqnarray}
(for $\delta^{-1}\equiv \delta$  and $l=0,...,k-1$, where  $k= {\max}_{t} k^{(t)}$).
Third, we should take into account for influence of traceless $L^{(t)}_{ij}$ and Young-symmetry $T^{(t)}_{rs}$ constraints on the structure of vertices $\big|  V^{(q)}_c\rangle^{(m)_{(i)_q}}_{(\vec{s})_{(i)_q}}$, $\big|  {V}{}^{(q)l}_c\rangle^{(m)_{(i)_q}}_{(\vec{s})_{(i)_q}}$ for $q=3,4,...,e$.

The resolution of    (\ref{g1})  for cubic vertices leads to the system
\begin{eqnarray}
\hspace{-0.7em}&\hspace{-0.7em}&\hspace{-0.7em} \mathcal{Q}(V^{(3)l-1}_{c|{(i)_3}},{V}^{(3)l}_{c|{(i)_3}}) = \sum_{n=1}^3
Q^{(i_n)} \big|  {V}^{(3)l-1}_c\rangle
   +  Q^{(i_t)}\Big( \big|  V^{(3)l-1}_c\rangle - \big|  {V}^{(3)l}_c\rangle\Big)=0
\label{g1operV3}   
\end{eqnarray}
 (for $t=1,2,3$  and  $l=0,...,k-1$) which particular solution for coinciding $V^{(3)}_{c|{(i)_3}} = {V}^{(3)0}_{c|{(i)_3}}=...={V}^{(3)k-1}_{c|{(i)_3}}$ has the usual form
but augmented  by the validity of spin,  traceless and Young-symmetry conditions
\begin{eqnarray}\label{gencubBRSTc}
 &&  Q_c^{tot}
\big|  V^{(3)}_c\rangle^{(m)_{(i)_3}}_{(\vec{s}_k)_{(i)_3}} =0,   \qquad    \big( L^{(t)}_{ij},\, T^{(t)}_{rs}\big) \big|  V^{(3)}_c\rangle^{(m)_{(i)_3}}_{(\vec{s}_k)_{(i)_3}} \ =\ 0, \\
 && \sigma^{(t)i_t}_c\big|  V^{(3)}_c\rangle^{(m)_{(i)_3}}_{(\vec{s}_k)_{(i)_3}}\ =\  \Big(s_{i_t}+\frac{d-2+\theta_{m_i,0}}{2}\Big)\big|  V^{(3)}_c\rangle^{(m)_{(i)_3}}_{(\vec{s}_k)_{(i)_3}} .
\label{gencubBRSTc1}
\end{eqnarray} where $ Q_c^{tot}=\sum_{t=1}^p Q_c^{(t)}$. The vertex $ \big| V^{(3)}_c\rangle^{(m)_{(i)_3}}_{(\vec{s}_k)_{(i)_3}} $ has a local representation:
\begin{equation}\label{xdep}
  \big |V^{(3)}_c\rangle^{(m)_{(i)_3}}_{(\vec{s}_k)_{(i)_3}} = \prod_{l=1}^3 \delta^{(d)}\big(x -  x_{i_l}\big) V^{(3)(m)_{(i)_3}}_{c|(\vec{s}_k)_{(i)_3}}(x)
  \prod_{l=1}^3 \eta^{(i_l)}_0 |0\rangle , \ \  |0\rangle\equiv \otimes_{t=1}^p |0\rangle^{t}.
\end{equation}.

Let us verify  that the deformed  equations of motion   and  any representative $\big|\widetilde{\chi}^{(i)}_c\rangle_{s_i}$ from arbitrary  gauge orbit $\mathcal{O}_{[1]|\chi^{(t)}_c}$ for field  and $\mathcal{O}_{[1]|\Lambda^{(t)l}_c}$ for $l$-th gauge parameter, for $ t=1,...,p $:
\begin{eqnarray}\label{goc1}
\hspace{-0.5em}&\hspace{-0.5em}&\hspace{-0.5em}\mathcal{O}_{[1]|\chi^{(t)}_c} = \big\{\big|\widetilde{\chi}^{(t)}_c\rangle_{\vec{s}_{k_t}} \big|\ \ \big|\widetilde{\chi}^{(t)}_c\rangle_{\vec{s}_{k_t}} =\big|{\chi}^{(t)}_c\rangle_{\vec{s}_{k_t}}  + \delta_{[1]}|{\chi}^{(t)}_c\rangle_{\vec{s}_{k_t}} ,  \, \forall \big|\Lambda^{(t)}_c\rangle_{\vec{s}_{k_t}}\big\}\\
\hspace{-0.5em}&\hspace{-0.5em}& \hspace{-0.5em} \mathcal{O}_{[1]|\Lambda^{(t)l}_c} = \big\{\big|\widetilde{\Lambda}^{(t)l}_c\rangle_{\vec{s}_{k_t}} \big|\ \ \big|\widetilde{\Lambda}^{(t)l}_c\rangle_{\vec{s}_{k_t}} =\big|{\Lambda}^{(t)l}_c\rangle_{\vec{s}_{k_t}}  + \delta_{[1]}|{\Lambda}^{(t)l}_c\rangle_{\vec{s}_{k_t}} ,  \, \forall \big|\Lambda^{(t)l}_c\rangle_{\vec{s}_{k_t}}\big\}, \label{goc2}
\end{eqnarray}
 (with  $(\epsilon, gh)\big|{V}{}^{(3)}\rangle = (1,3)$)
  for interacting fields remains by traceless and mixed-symmetric
after applying the deformed gauge transformations. It is sufficient to find that for any  constraint $A^{(t)}\in \{L^{(t)}_{ij}, T^{(t)}_{rs} \}$ one should be
\begin{eqnarray}
&\hspace{-0.5em} &\hspace{-0.5em}  A^{(t)}\delta_{[1]}|{\Lambda}^{(t)l-1}_c\rangle_{\vec{s}_{k_t}}= A^{(t)}Q_c^{(t)}|\Lambda^{(t)l}_c\rangle_{\vec{s}_{k_t}} - g \int d\eta^{(i_1)}_0 d\eta^{(i_2)}_0\Big( {}_{\vec{s}_{k_{i_1}}}\langle
\Lambda^{(\{i_1)l}_c\big| \nonumber\\
   &\hspace{-0.5em} &\hspace{-0.5em}
\ \   \phantom{\delta_{[1]} \big| \chi^{(i)} \rangle_{s_i}}
\otimes{}_{\vec{s}_{k_{i_2}}}
   \langle \chi^{(i_2\})}_c\big| A^{(t)}\big|{V}{}^{(3)}_c\rangle^{(m)_{i_2t}}_{(\vec{s}_k)_{(i)_2t}} = 0, \label{cubgtrcc} \\
  &\hspace{-0.5em} &\hspace{-0.5em}  A^{(t)}\frac{\overrightarrow{\delta} S^{(m)_3}_{[1]C|(\vec{s})_3}}{\delta {}_{\vec{s}_{k_t}}
   \langle \chi^{({t})}_c\big|}  =A^{(t)}Q_c^{(t)}|\chi^{(t)}_c\rangle_{\vec{s}_{k_t}} + g
    \sum_{1\leq i_1<i_2\leq p}^{i_j\ne t} \hspace{-1.0em}
   \int  \prod_{j=1}^2 d\eta^{(i_j)}_0   {}_{\vec{s}_{k_{i_j}}}\langle \chi^{(i_j)}_c \big|\nonumber\\
   &\hspace{-0.5em} &\hspace{-0.5em}
\ \   \phantom{\delta_{[1]} \big| \chi^{(i)} \rangle_{s_i}}\otimes
 A^{(t)}\big|  V^{(3)}_c\rangle^{(m)_{(i)_2t}}_{(\vec{s}_k)_{(i)_2t}}= 0. \label{cubgtr1c}
\end{eqnarray}

We have shown, that imposing of  traceless and Young symmetry constraints on fields and gauge parameters (\ref{L11T})  represents the necessary but not sufficient condition for the consistency
of cubically  deformed  Lagrangian dynamics.
Indeed, in this case the latter terms in (\ref{cubgtrcc}) and (\ref{cubgtr1c}) do not vanish. and therefore the number of independent initial data (number of p.d.f.) for deformed and free cases become different.

Obvious generalization of this requirement for the $q$-tic vertices  leads for $\big|  {V}^{(q)l}_c\rangle^{(m)_{(i)_q}}_{(\vec{s}_k)_{(i)_q}}$ (i.e. for coinciding vertices) in addition to the rest equations from (\ref{g2}), (\ref{ge}) to be annihilated  by constraints $L^{(t)}_{ij}$, $T^{(t)}_{rs}$.
\begin{equation}\label{ptrace}
  \big(L^{(t)}_{ij}, \, T^{(t)}_{rs}\big)\big|  {V}^{(q)l}_c\rangle^{(m)_{(i)_q}}_{(\vec{s}_k)_{(i)_q}}\ =\ 0,\ \ q=3,4,\ldots , e; \  l=-1,0,1,...,k-1.
\end{equation}
  The $Q_c^{tot}$-closed as well as  traceless- and Young-symmetric    solutions for the  equations  (\ref{gencubBRSTc}), (\ref{gencubBRSTc1})
determines consistent cubic vertices for irreducible interacting  mixed-symmetric HS fields with given masses and spins which LF has  the same number of p.d.f. as ones for the same free irreducible fields. The cubic vertices in \cite{BRST-BV3} do not satisfy to this property.

It is not difficult  to find  $L^{(t)}_{ij}$-traceless and $T^{(t)}_{rs}$-mixed-symmetric  solution  $$\big|  \overline{V}{}^{(3)}_c\rangle^{(m)_{(i)_3}}_{(\vec{s}_k)_{(i)_3}} \equiv \big|  V^{(3)}_{c|{irrep}}\rangle^{(m)_{(i)_3}}_{(\vec{s}_k)_{(i)_3}}$$
of the equations (\ref{gencubBRSTc}), (\ref{gencubBRSTc1})   with  $Q_c^{tot}$-closed  vertex  $\big|  V^{(3)}_{c|{irrep}}\rangle^{(m)_{(i)_3}}_{(\vec{s}_k)_{(i)_3}}$   as follows (e..g.  for $p=3$ for the $Q_c$-closed vertex $\big|  V^{(3)}_c\rangle^{(m)_3}_{(\vec{s}_k)_{(3}}$ from \cite{BRST-BV3})
\begin{eqnarray}\label{L11V+}
 \hspace{-0.5em}&\hspace{-0.5em}&\hspace{-0.5em}  \big| \overline{ V}{}^{(3)}_c\rangle^{(m)_3}_{(\vec{s}_k)_{3}} = \ \prod^3_{t=1}\prod^{k_t}_{i_t=1}\mathbf{P}^{(t|L)}_{i_ti_t}\prod^{k_t}_{1\geq i_t<j_t<k_t}\mathbf{P}^{(t|L)}_{i_t j_t}\prod^{k_t}_{1\geq r_t<s_t<k_t}\mathbf{P}^{(t|T)}_{r_t q_t}\big|  V^{(3)}_c\rangle^{(m)_3}_{(\vec{s}_k)_{(3}}.
 \end{eqnarray}
 Here, operators $\mathbf{P}^{(t|L)}_{i_ti_t}$, $\mathbf{P}^{(t|L)}_{i_t j_t}$ and $\mathbf{P}^{(t|T)}_{r_t q_t}$ are respectively projectors for fixed  $t$ on pure traceless modes with respect to indices in $i_t$ group, mixed traceless modes with respect to indices in $i_t$ and $j_t$ groups and Young-symmetry modes with respect to indices in $r_t$ and  $q_t$ groups.
 Explicitly,
 \begin{eqnarray}
  \hspace{-1.2em} &\hspace{-1.2em}& \hspace{-1.3em} \mathbf{P}^{(t|L)}_{i_ti_t} \hspace{-0.3em}  =  \hspace{-0.2em} \sum_{j=0}^{[s_{i_t}/2]}(-1)^j\frac{C(s_{i_t}-1-j,d/2)}{j!{}C(s_{i_t}-1,d/2)}(L^{(t)+}_{i_ti_t})^j (L^{(t)}_{i_ti_t})^j,  \  C(i,j)\equiv \prod_{l=0}^{i-1}(l+j),\nonumber \\
    &\hspace{-1.2em}& \hspace{-1.3em}  \mathbf{P}^{(t|L)}_{i_tj_t}   =  \sum_{j=0}^{\min([s_{i_t}/2], [s_{j_t}/2])}(-1)^j\frac{4^j{}C(s_{i_t}+s_{j_t}-1-j,d)}{j!{}C(s_{i_t}+s_{j_t}-1,d)}(L^{(t)+}_{i_tj_t})^j (L^{(t)}_{i_tj_t})^j,  \label{L11V+w}\\
   && \mathbf{P}^{(t|T)}_{r_tq_t}   =  \sum_{j=0}^{ s_{q_t}}(-1)^j\frac{(s_{r_t}-s_{q_t}+1-j)!}{j!{}(s_{r_t}-s_{q_t}+1)!}(T^{(t)+}_{r_tq_t})^j (T^{(t)}_{r_tq_t})^j,  \nonumber .
\end{eqnarray}
Substituting of found cubic vertex $\big| \overline{ V}{}^{(3)}_c\rangle$ (\ref{L11V+}) into  the action (\ref{S[e]}), (\ref{S[3]})  and sequence of gauge transformations  (\ref{gt1}),  leads to the same properties (with preserved p.d.f.) of the LF for cubically  interacting HS fields with given spins as ones for undeformed  model for $p$-samples of free fields.
Note, the above   way to get traceless and Young-symmetric cubic vertex can be immediately applied for the quartic and for $e$-ctic, $e \geq 4$  vertices.

\section{Conclusion}
\label{sec:conclusion}

In the present article, it is derived  the criteria  for cubic and $p$-tic ($p\geq 4$) vertices
obtained within approach with incomplete BRST operator to describe
consistent Lagrangian dynamics for irreducible interacting
mixed-symmetric HS fields in $d$-dimensional Minkowski
space subject to appropriate holonomic constraints. It is shown that
imposing of only constraints above on field and gauge parameter
vectors (\ref{L11T}) that form gauge-invariant content of  LF is insufficient to preserve the number of p.d.f. passing from free to interacting theory.
Additionally,  to above restrictions, the set of incomplete BRST,
spin operators,  cubic vertices and holonomic constraints must form
closed superalgebra (\ref{QsL11}), (\ref{gencubBRSTc}),
(\ref{gencubBRSTc1}). The constraints and cubic (also all $p$-tic)
vertices should supercommute. The solution for BRST-closed
traceless, Young-symmetric cubic vertices represents for constrained case  the
local three-vector. At the same time  the application of the
approach with complete BRST operator as we expect   automatically
leads to local cubic vertices to be by BRST ($Q^{tot}$)-closed solution with given
spin  with the same interacting
HS fields, but depending  on  wider set of oscillators with
additional trace- and Young-inspired factors. The
correspondence among the cubic vertices in   approaches with complete and incomplete BRST operators  for the same irreducible  interacting
fields may be established according to the  receipt for totally-symmetric HS fields \cite{Rcubmasless},  \cite{BRcubmass}, as well as in BRST-BV approach  according to \cite{RBRSTBV}.
\paragraph{Acknowledgements} The author is thankful to organizers and participants of International
Workshop "Supersymmetries and Quantum Symmetries 2024" for
hospitality and illuminating discussions of the results.
%


\end{document}